\documentclass[3p,times,procedia]{elsarticle}
\usepackage{nupha_ecrc}
\usepackage{wrapfig}
\usepackage{amsmath}
\usepackage{bm}
\usepackage{color}

\volume{00}

\firstpage{1}

\journalname{Nuclear Physics A}

\runauth{}

\jid{nupha}

\jnltitlelogo{Nuclear Physics A}

\usepackage{amssymb}

\usepackage[figuresright]{rotating}

\begin{document}

\begin{frontmatter}



\dochead{}

\title{Does non-monotonic behavior of directed flow signal
the onset of deconfinement ?}

\author[label1]{Yasushi Nara}
\author[label2]{Akira Ohnishi}

\address[label1]{Akita International University,
Yuwa, Akita-city 010-1292, Japan}
\address[label2]{Yukawa Institute for Theoretical Physics,
Kyoto University, Kyoto 606-8502, Japan}

\begin{abstract}
We investigate the effects of nuclear mean-field
as well as the formation and decay of nuclear clusters
on the directed flow
$v_1$ in high energy nucleus-nucleus collisions
from $\sqrt{s_{NN}}=7.7$  GeV to 27 GeV incident energies
within a transport model.
Specifically, we use the JAM transport model in which
potentials are implemented based on the framework of
the relativistic quantum molecular dynamics.
Our approach reproduces the rapidity dependence of directed flow data up to
$\sqrt{s_{NN}}\approx 8$ GeV showing the significant importance of mean-field.
However, the slopes of $dv_1/dy$ at mid-rapidity are calculated to be positive
at $\sqrt{s_{NN}}=11.7$ and 19.6 GeV, and becomes negative above 27 GeV.
Thus the result from the JAM hadronic transport model with nuclear mean-field approach
is incompatible with the data.
Therefore within our approach, we conclude that the excitation function of
the directed flow cannot be explained by the hadronic degree of freedom alone.
\end{abstract}


\end{frontmatter}


\section{Introduction}
\label{sec:intro}

Determination of the equation of state (EoS) at high density QCD
matter from an anisotropic flow in heavy ion collisions
has been discussed for a long time.
In particular, the softening of the EoS influences 
drastically the nucleon directed flow, and collapse of the directed flow
$v_1=\langle \cos\phi\rangle$
has been suggested as a signal of the phase transition from
ordinary hadronic matter to quark-gluon plasma (QGP) 
~\cite{Rischke:1995pe,Stoecker:2004qu}.
The slope of nucleon $v_1$ is normally positive in the hadronic scenario, but
hydrodynamical calculations with QGP EoS
predict a negative slope of $v_1$ at mid-rapidity,
when matter passes through the softest point of the EoS
~\cite{Csernai:1999nf,Brachmann:1999xt}.
On the other hand, microscopic hadronic transport calculations also
yield a negative slope due to geometrical effects at sufficiently
high collision energies~\cite{Snellings:1999bt,Bleicher:2000sx}.
The theoretical studies on the beam energy dependence of
the directed flow based on the newly developed models
such as hybrid transport approach~\cite{Steinheimer:2014pfa},
three fluid model~\cite{Ivanov:2014ioa}, and
the PHDS transport model~\cite{Konchakovski:2014gda}
have been extensively performed.
However, a definite conclusion has not been drawn so far as to the
interpretation of the directed flow data measured by the STAR collaboration
~\cite{Adamczyk:2014ipa}.

In this work, we compute beam energy dependence of the directed flow
in the energy range of the beam energy scan (BES) program at RHIC
within a hadronic transport model in which
baryon mean-field is implemented within the formalism of
the simplified version of the relativistic quantum molecular dynamics.
The effects of nuclear cluster formations
and their statistical decay on the spectra are also investigated.

\section{Hadronic trasport model}
\label{sec:model}

We employ a hadronic transport model JAM
that has been developed
based on the resonance and string degrees of freedom~\cite{Nara:1999dz}.
Particle productions are modeled by the resonance or string excitations
and their decays.
Secondary interactions among produced particles are also included
via the two-body collision.
Nuclear mean-field of baryons are
included based on the framework of a simplified version of
relativistic quantum molecular dynamics (RQMD/S)~\cite{Isse:2005nk}.
We adopt the following 
Skyrme-type density dependent and Lorentzian-type momentum dependent
mean field potential for baryons,
\begin{equation}
U(\bm{r},\bm{p})=
\alpha \left(\frac{\rho(\bm{r})}{\rho_0}\right)
+\beta \left(\frac{\rho(\bm{r})}{\rho_0}\right)^\gamma
+\sum_k \frac{C_k}{\rho_0} \int d\bm{p}'
\frac{f(\bm{r},\bm{p}')}{1+\left[(\bm{p}-\bm{p}')/\mu_k\right]^2}
\ ,
\end{equation}
where $f(\bm{r},\bm{p})$ is the phase space distribution function, and its
integral over $\bm{p}$ becomes the density distribution $\rho(\bm{r})$.
In RQMD/S, $\rho(\bm{r})$ is computed by assuming Gaussian distribution
function.
We use the parameter set which yields the incompressibility of $K=272$ MeV;
$\alpha=-0.209$ GeV, $\beta=0.284$ GeV, $\gamma=7/6$,
$\mu_1=2.02$ fm$^{-1}$, $\mu_2=1.0$ fm$^{-1}$,
$C_1=-0.383$ GeV, $C_2=0.337$ GeV, and $\rho_0=0.168$ fm$^{-3}$.

The formation of nuclear clusters are taken into account
based on the phase space distribution
of nucleons at the end of the simulation.
If nucleons are close in the phase space, 
nuclear cluster is formed:
if the relative distances and momenta between nucleons are less than
$R_0$ and $P_0$ at the same time, these nucleons are considered to 
belong to the same nuclear cluster.
Coalescence parameters $R_0=4.0$ fm and $P_0=0.3$ GeV/c are chosen
by fitting the proton rapidity distribution at bombarding energy of
$\sqrt{s_{NN}}=2.7$ GeV for central Au+Au collisions.
This parameter set gives fairly good description of 
the rapidity distribution of protons for a wide range of
collision energies.
Nuclear clusters are generally not in their ground states,
but in excited states. Thus the statistical
decay of such excited fragments are also
taken into account~\cite{Maruyama:1992bh}.
In the statistical decay model (SDM), we include the emissions
of nuclei up to the mass number of 4 as well as gamma emission.

\section{Results}
\label{sec:result}

\begin{wrapfigure}[23]{l}{7.0cm}
\vspace*{-\intextsep} %
\includegraphics[width=7.0cm]{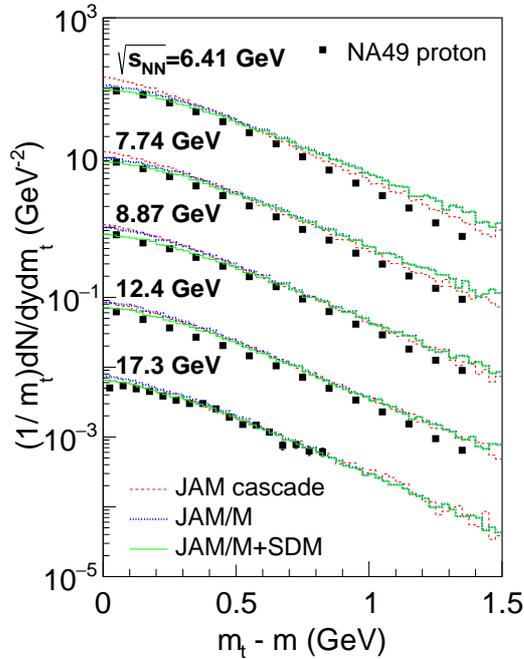}
\caption{Transverse mass distributions for protons 
in central Pb+Pb collisions at 
$\sqrt{s_{NN}}=6.41-17.3$ GeV~\cite{Alt:2006dk}
is compared to JAM cascade mode (dashed lines),
and JAM/M (dotted lines), and JAM/M+SDM (solid lines).
}
\label{fig:na49pmt}
\end{wrapfigure}

In Fig.~\ref{fig:na49pmt}, we compare the transverse mass distributions
of protons in central Pb+Pb collisions at
$\sqrt{s_{NN}}=6.41-17.3$ GeV from NA49~\cite{Alt:2006dk}
with the JAM results.
The spectra are scaled down by successive factors of 10
from the 6.41 GeV data.
The proton distributions from JAM cascade mode (dashed lines)
overestimate the yield at low transverse mass region.
It is seen that 
JAM mean-field calculation (JAM/M) suppresses the
yields of the low transverse momentum except the highest NA49 energy,
but still predicted yields are slightly higher than the data.

It is found that the proton stopping can be improved by
taking into account nuclear cluster formations~\cite{Li:2015pta},
which contributes to about 20\% reduction of the proton rapidity
distribution.
We also found the similar results which affect the transverse mass
distribution in the
low transverse mass region.
Inclusion of nuclear cluster formation improves the description
of the proton transverse mass distribution as shown in
the Fig.\ref{fig:na49pmt} (JAM/M+SDM).
In general, nuclear fragments are in the excited states and
decay by emitting particles.
Thus their statistical decay into the ground state should be considered.
It turns out that cluster formation affects the
low transverse mass region, but the contribution of
the statistical decay is negligibly small in the NA49 energy ranges.
We found that the statistical decay of nuclear cluster is only
important at lowest AGS energy.
The effect of the mean-field at higher transverse mass region
is to harden the spectra  at lower collision energies of
6.41 and 7.74 GeV.

\begin{wrapfigure}[47]{l}{7.0cm} \vspace*{-\intextsep} %
\includegraphics[width=8.0cm]{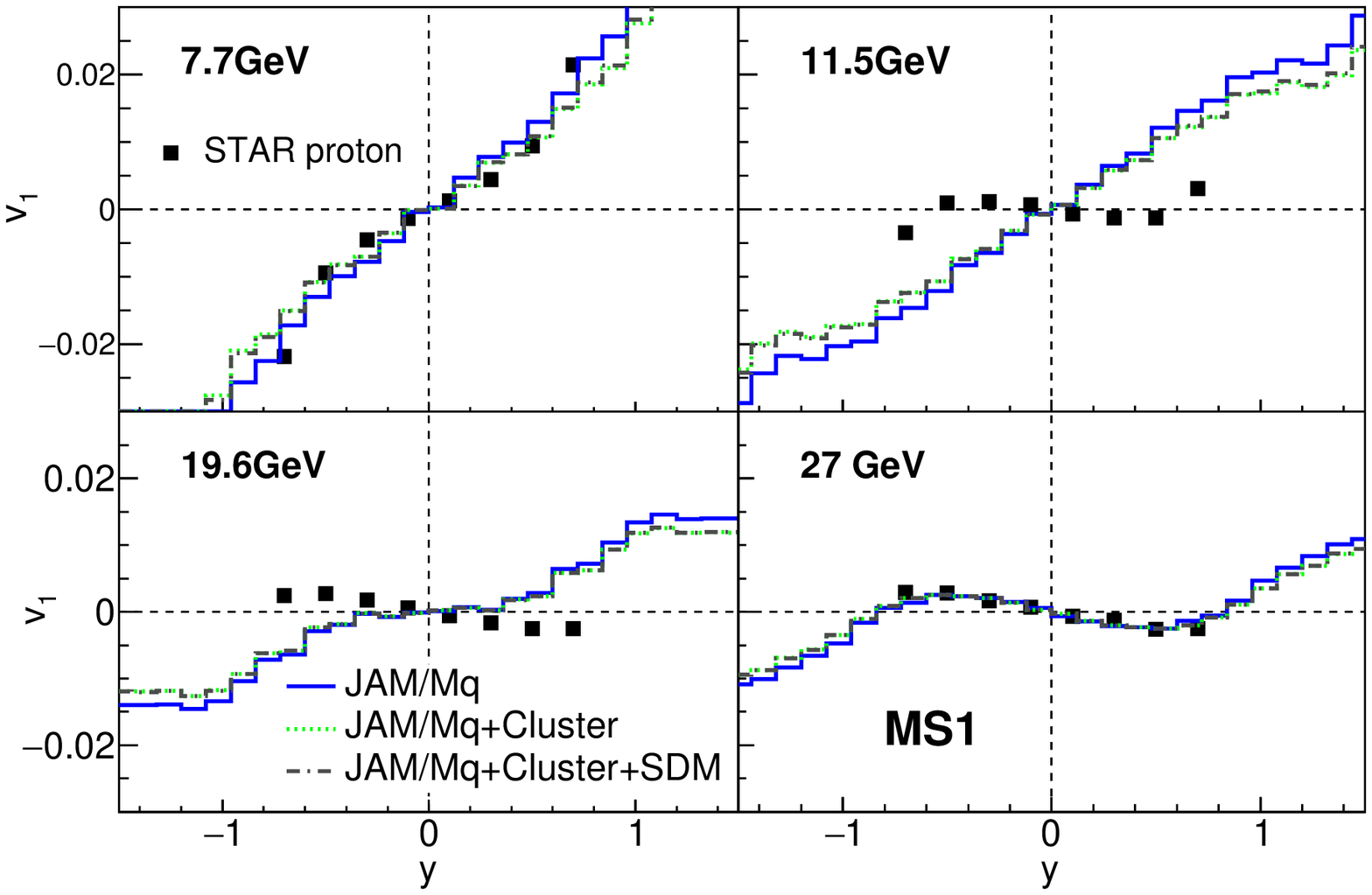} \caption{Directed flows of protons
in mid-central Au+Au collisions (10-40\%) at $\sqrt{s_{NN}}=7.7-27$ GeV from
JAM mean-field mode (dashed lines), and JAM mean-field followed by the
statistical decay (solid lines)
in comparison with the STAR data~\cite{Adamczyk:2014ipa}.  }
\label{fig:starv1c} \includegraphics[width=7.5cm]{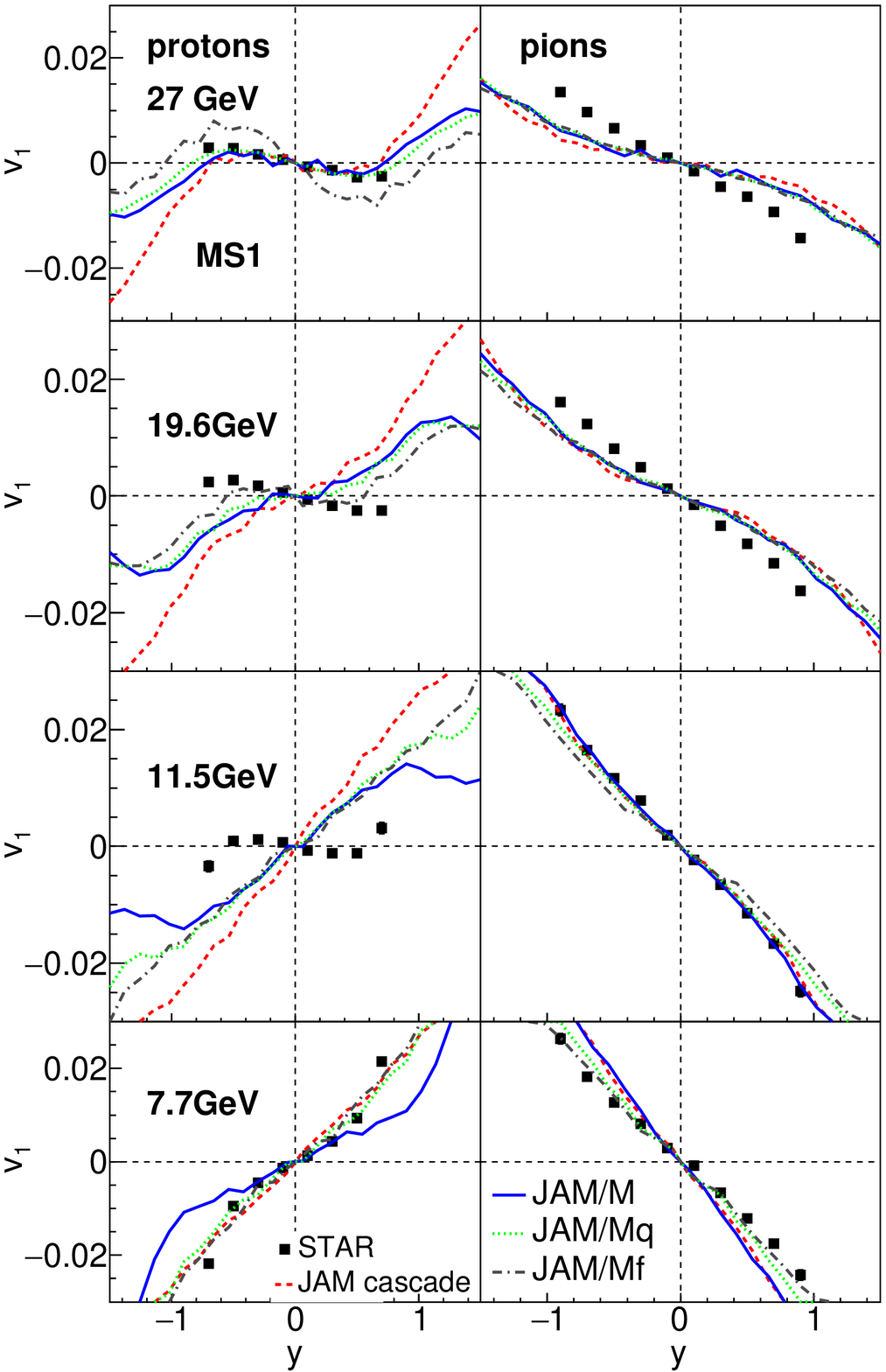}
\caption{Directed flows of protons and pions in mid-central Au+Au collisions
(10-40\%) at $\sqrt{s_{NN}}=7.7-27$ GeV from JAM cascade mode (dashed lines),
and JAM cascade with attractive orbit (solid lines)
in comparison with the STAR data~\cite{Adamczyk:2014ipa}.  }
\label{fig:starv1mid} \end{wrapfigure}

Let us now discuss the potential effects and formation
of nuclear clusters on the directed flow.
First we study the effect of nuclear cluster formation and its decay
on the $v_1$.
In Fig.~\ref{fig:starv1c},
we compare proton directed flow to
the STAR data for mid-central Au+Au collisions~\cite{Adamczyk:2014ipa}.
It is seen that the effect of nuclear cluster formation
on the proton $v_1$
 is about 15\%, and the effect of the statistical
decay of nuclear cluster is very tiny.
Thus we conclude that formation of nuclear cluster and its decay
plays no role in the proton $v_1$ at mid-rapidity
from the viewpoint of the softening of the EoS.

Figure~\ref{fig:starv1mid} shows the rapidity dependence of the
proton and pion directed flow for mid-central Au+Au collisions
at $\sqrt{s_{NN}}=7.7, 11.5, 19.6$ and 27 GeV.
JAM cascade agree with the 7.7 GeV data, but
JAM cascade predictions does not show negative
slopes at both 11.5 and 19.6 GeV, which disagree with the data.
The negative slope of $v_1$ at 27 GeV may be due to a geometrical effect
as discussed in Refs.~\cite{Snellings:1999bt,Bleicher:2000sx}.

We tested three different implementations of the baryon potentials.
In the JAM/M model, potentials are only included for the formed
baryons (solid lines).
String excitation dominates the particle productions
in the energy ranges considered here,
and there are many hadrons which are not formed during a formation
time (pre-formed hadrons).
However, leading hadrons normally contain original constituent
quarks, and they may interact with other hadrons.
Interaction of constituent quarks are included effectively
in the two-body collision term by reducing the hadronic cross sections
~\cite{Bass:1998ca}.
We apply the same idea for the mean field part (JAM/Mq): hadrons which
has original constituent (di)quarks interact by reducing
the strength of the potential by (2/3) 1/3 during their formation time.
The results from the model JAM/Mq are plotted in the dotted lines.
Finally, we also display the results of JAM/Mf in which
all hadrons including pre-formed hadrons fully feel potentials,
in order to see somewhat maximum effects of the baryon potentials.
From the results obtained by three scenarios, it is seen that
the effects of hadronic mean-field is large, and to reduce the slope of
the proton directed flow, but still incompatible to the data.
Namely, our model predict positive $dv_1/dy$ at 11.5 and 19.6GeV.
On the other hand, pion directed flow is not largely affected
by the mean-field, since the meson is affected indirectly from
baryons, not from meson mean-field in the current model.

\section{Conclusion}
\label{sec:concl}

In conclusion, we have examined the effects of baryon mean-field potentials
as well as the nuclear cluster formations and their statistical decays
on the directed flow at BES energies by the hadronic transport model JAM.
The Skyrme-type density dependent and Lorentzian-type momentum dependent
mean field potentials
are implemented within a framework of RQMD/S.
We found that the baryon mean-field
reduces the slope of $dv_1/dy$ by 20-30 \%.
The effect of nuclear cluster formation on the proton
spectra as well as $v_1$ is also found to be about 10-20\%.
Contributions from the statistical decay of nuclear fragment
is very small.
We tested three different implementations of potentials.
All of them cannot explain the correct beam energy dependence
of the proton directed flow especially at
the transition region of the reverse of the $dv_1/dy$ 
around $9 \lessapprox \sqrt{s_{NN}}\lessapprox 20$ GeV.
We finally remark that the negative $dv_1/dy$ at 11.5 and 19.6 GeV
can be described by the transport approach
if the effects of the softening of the EoS is taken into account
~\cite{NOS:2016}.

\section*{Acknowledgement}
This work was supported in part by
the Grants-in-Aid for Scientific Research from JSPS
(Nos.
    15K05079, 
    15K05098
    ),
    the Grants-in-Aid for Scientific Research on Innovative Areas from MEXT
     (
      Nos. 24105001, 24105008),
       by the Yukawa International Program for Quark-Hadron Sciences,


\bibliographystyle{elsarticle-num}
\bibliography{qm15}


\end{document}